\pdfoutput=1
\documentclass[aps,pre, twocolumn, groupedaddress]{revtex4-1} 
\usepackage{lipsum}
\usepackage{mathtools}
\usepackage{graphicx}
\usepackage{dcolumn}
\usepackage{amsmath}    % need for subequations
\usepackage{amssymb}
\usepackage{bm}
\usepackage{hyperref}
\usepackage{latexsym}
\usepackage{verbatim}
\usepackage{color}
\usepackage[caption=false]{subfig}

\def\beq{\begin{equation}}
\def\eeq{\end{equation}}

\def\beq{\begin{equation}}                          
\def\eeq{\end{equation}}                          
\def\bea{\begin{eqnarray}}                          
\def\eea{\end{eqnarray}}

\DeclareRobustCommand{\uvec}[1]{{%
  \ifcsname uvec#1\endcsname
     \csname uvec#1\endcsname
   \else
    \bm{\hat{\mathbf{#1}}}%
   \fi
}}

\preprint{}

\begin{document}

%%%%%%%%%%%%%%%%%%%%%%%%%%%%%%%%%%%%%%%%%%%%%%%%%%%
%                               TITLE & ABSTRACT
%%%%%%%%%%%%%%%%%%%%%%%%%%%%%%%%%%%%%%%%%%%%%%%%%%%
%Title of paper
\title{Dynamics of active run and tumble and  passive particles in binary mixture}
\author{Vivek Semwal}
\email{viveksemwal.rs.phy17@itbhu.ac.in}
\affiliation{Indian Institute of Technology (BHU) Varanasi, India 221005}
\author{Jay Prakash Singh}
\email{jayps.rs.phy16@itbhu.ac.in}
\affiliation{Indian Institute of Technology (BHU) Varanasi, India 221005}
\author{Shradha Mishra}
\email[]{smishra.phy@itbhu.ac.in}
\affiliation{Indian Institute of Technology (BHU) Varanasi, India 221005}
\date{\today}
\begin{abstract}
	We  study  a binary mixture of disk-shaped {\it active run and tumble } particles (${ARNPs}$) and
passive particles on a two-dimensional substrate. Both types of particles are athermal. The
particles interact through the soft repulsive potential. The activity of $ARNPs$
is controlled by tuning their tumbling rate.  The system is studied for various sizes of passive particles keeping  size of $ARNPs$ fixed. Hence the variables are, size ratio (S) of passive particles and $ARNPs$, and activity of $ARNPs$ is $v$. The characterstics dynamics of both $ARNPs$ and passive particles 
	show a crossover from early time ballistic to later time diffusive. Furthermore, we observed that passive particles dynamics changes from diffusive 
	to subdiffusive  with respect to their size. Moreover, late time effective diffusivity $D_{eff}$ of passive particles 
	decreases with increasing their size as in the corresponding  equilibrium  Stokes systems. We calculated  the effective tempratures, using $D_{eff}$, $T_{a,eff}(\Delta)$ and also using speed distribution $T_{a,eff}(v)$ and compared them. The  both  $T_{a,eff}(\Delta)$ and $T_{a,eff}(v)$  increases linearly with activity and are in agreement with each other. Hence we can say that an effective equilibrium can be establish in such binary mixture.
Our study can be useful to study the various biological systems like; dynamics of passive organels in cytoplasm, colloids etc.
	
\end{abstract}
\maketitle
\section{Introduction}
In the recent years, researchers have paid a lots of attention in the field of active matter \citep{sriram2005, viscek2012,marchettirmp, sriram,cates2015,cates2013} because of their unusual properties in comparison to their equilibrium counterparts. Examples of active systems ranges from microscale such as bacterial colonies, cell suspension, artificially designed microparticles \citep{sizeratio,vivek,janus,janus2,janus3,janus4}, etc. to the larger scale; fish school, flock of birds {\cite{flock,fish,sriramflock}} etc. 
 Active system continuousely evolve with time which leads to nonequilibrium class with intresting features i.e; pattern formation \cite{patternform}, nonequilibirum phase transition \cite{phasetransition, sudip2018, biplab, jp2021},
large density flcutuations \cite{density}, enhanced dynamics
{\cite{enhance1,enhanced, vivek}}, motility induced phase separation \cite{mipsc,phases, mips2013prl, bechinger} etc.
In recent years,  the motion of passive particles 
in the presence of an active medium
is used to explore the nonequilibrium properties of the medium. 
In such  mixtures passive  particles  exhibit enhanced diffusivities $D_{eff}$ greater than their thermal (Brownian) diffusivity $D_0$ \cite{en2,enhance3}. 
In the experiment of \cite{wu}, passive Brownian disks in active bacterial solution show enhanced diffusivity. 
The enhanced diffusivity $D_{eff}$ increases linearly with increasing concentration of bacteria in the solution \cite{enhancediffusivity,enhancediff,enhce}.
A variety of studies have focused on the role of bacterial concentration on passive particle diffusion.
The effect of passive particles size is still not clear. In the absence of bacteria, or the equilibrium fluid the diffusivity of a sphere follows the Stokes-Einstein relation \cite{stoke}.
To understand the role of particle size on their dynamics in the active medium,
 we introduce a binary mixture 
of active run and tumble $ARNPs$ and passive particles.
$ARNPs$ move in a straight line for some time and then undergo a random rotation (tumble event).
%.are characterised by their characterstic movement which includes persistent motion followed by diffusion or random movement.}}.
Hence activity can also be tuned with tumbling rate. A large tumbling rate means smaller run time and hence more random motion.\\
 We studied  the mixture for different sizes of passive particles and the activity of $ARNPs$. 
	Effective diffusity $D_{eff}$ of $ARNPs$ does not change significantly whereas it decreases linearly with size for the passive particles: similar to their equilibrium counterparts-Stokes system \cite{stoke}. 
	We  calculated  the effective temperatures, using $D_{eff}$, $T_{a,eff}(\Delta)$ and using 
	speed distribution of $ARNPs$, $T_{a,eff}(v)$.
	%\textcolor{red}{effective temperature via effective diffusivity and maxwell boltzmann distribution both are match well}and effective diffusivity $D_{eff}$ of $ARNPs$ with activity. 
	The both increases linearly with increasing activity for all size ratios.  
Hence although the system
        is active, an effective equilibrium can be established in such mixtures.\\

%	with arguing the stablishment of an effective equilibrium in such systems.\\ 
The rest of paper is organised as follows. In section \ref{secI} we discuss the model with simulation details. In section.\ref{Result} we discuss the results followed by conclusion in \ref{secdis} at the end.
\section{Model}\label{secI}
We consider  a binary mixture of $N_a$ small run and tumble particles $ARNPs$ of radius $r_a$, and $N_p$ passive particles 
of  radius $r_p$ moving on a two-dimensional (2D)  substrate of size $L \times L$. 
The size of the $ARNPs$ particle is kept 
fixed whereas the size of passive particles is tuned. We define the size ratio $S = r_{p}/ r_{a}$. The position 
vector of the centre of the $i^{th}$ $ARNPs$ and  passive particle at time $t$ is given by 
${\bf r}_{i}^{a}(t)$ and ${\bf r}_{i}^{p}(t)$, respectively. The orientation of $i^{th}$ $ARNPs$ is represented 
by a unit vector $ {\bf n}_i = (\cos{\theta_i},\sin{\theta_i})$. The dynamics of the 
$ARNPs$ particle is governed by the overdamped Langevin equation \cite{filly,langevin,langevin2}
\begin{equation}
	\partial_t{\bf{r}}_i^{a}=v_0{\bf{n}_i}+\mu_1\sum_{j\neq i}{\bf {F}}_{ij}
\label{eq(1)}
\end{equation}
% \begin{equation}
%	 \partial_t\theta_i={\eta}^r_i(t)
%\label{eq(2)}
%\end{equation}
The first term on the right-hand side (RHS) of Eq. \ref{eq(1)} is due to the activity of the $ARNPs$, and  $v_0$ is the self-propulsion speed. 
The second term, the force $F_{ij}$ is the soft repulsive interaction among the particles. 
It is obtained from the binary soft repulsive pair potential $V(r_{ij} ) =\frac{k(r_{ij}-2\sigma)^2}{2}$ and ${\bf F}_{ij} = -\nabla V(r_{ij})$,
for $r_{ij} \le  \sigma$ and zero otherwise. $\sigma=a_{i}+a_{j}$, where $a_{i,j}$ is the radius of $i^{th}$ and $j^{th}$ particle respectively.
 $r_{ij} = |r_{j}-r_{i}|$ is the distance between particle
$i$ and $j$. The summation runs over all the particles.
$\tau=(\mu k)^{-1}$ sets the elastic time scale in the system. 
Further, the orientation of $ARNPs$ is controlled by run and tumble events. The particles
orientation is updated by Eq. \ref{eqn:2} introducing a uniform random number $r_n$. A tumbling rate $\lambda$ is 
defined such that if $\lambda > r_n$ then the particle undergoes a tumble event with a random orientation $\eta_i \in (-\pi, +\pi)$.
Else it undergoes run event with the same angle as in previous step.
Hence large tumbling rate $\lambda$ means frequent change in particle orientation. Hence the orientation
update of $ARNPs$ is given by:

\begin{equation}
\label{eqn:2}
\theta_{i}(t+\Delta t)=\theta_{i}(t)+\eta_{i}(t)
\end{equation}
 
The position of the passive particles  is also governed by the overdamped Langevin equation, 
\begin{equation}
	\partial_t{\bf{r}}_i^{p}=\mu_2\sum_{j}{\bf {F}}_{ij}
	\label{eq(3)}
\end{equation}
The ${\bf F}_{ij}$ has the same form as defined in Eq. \ref{eq(1)}. 
There is no translational noise \cite{noise} in Eqs. \ref{eq(1)} and \ref{eq(3)}, therefore, both
$ARNPs$ and passive particles are athermal in nature.

The smallest time step 
considered is $\Delta t=5\times10^{-4}$, much smaller than the elastic time scale $\tau$  (for  $\mu=1$ and $k=1$)  The system is simulated for total time steps of $t=2\times10^{5}$. 
All the physical quantities calculated here are averaged over $50$ independent
realizations. The self-propulsion speed $v_0$ is kept fixed to $0.5$ and 
activity is varied by tuning tumbling rate $\lambda$, such that the 
dimensionless activity defined as $v=v_0/\lambda r_{a}$ can vary from $5\times10^3$ to 
$5\times10^4$. 
%\textcolor{red}{where dimensionless tumbling rate $v$ defined as $v=v_0/\lambda r_{a}$ where $v_0$ is self propulsion speed is 0.5 and $\lambda$ rate of tumble,$r_{a}$ is radius of active particle.} 
 
We start with random initial positions and orientation directions of all the particles. 
Once the update of the above two equations is done for all  $N=N_{a}+N_{p}$ particles, it is counted as one simulation step.  We simulated the system for total $2 \times 10^5$ steps. 
Packing fractions of $ARNPs$ and passive particles is fixed  $\frac{\pi( N_{a}r_{a}^{2}+N_{p}r_{p}^2)}{L^{2}}=0.58$. The linear dimensions of the system is fixed to $L = 150r_a$.

\section{Results \label{Result}}
\subsection{Dynamics of the particles in the mixture}
\begin{figure} 
\label{fig:1}
\centering
{\includegraphics[width=0.9 \linewidth]{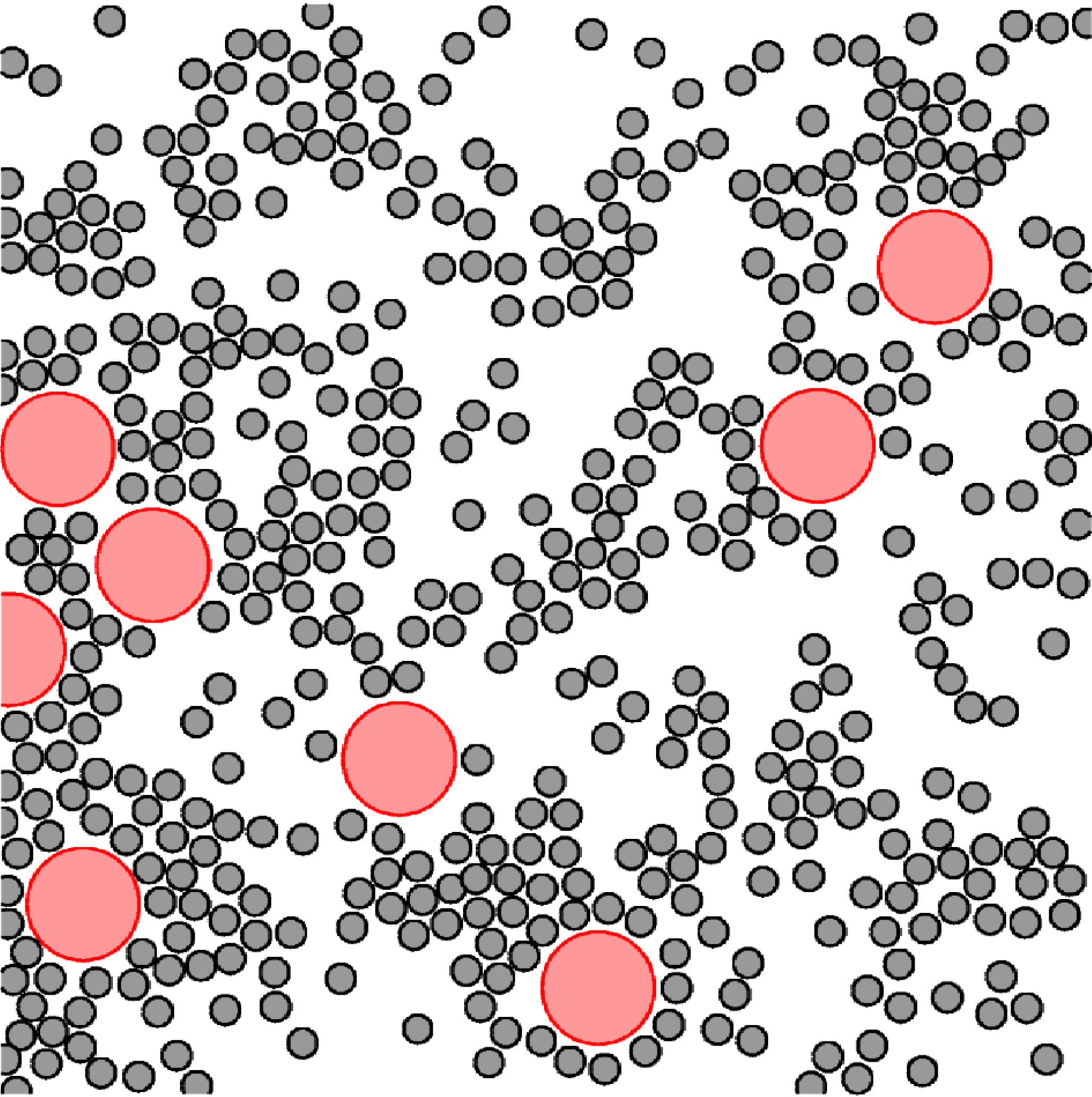}}
	\caption{(color online) Typical snapshot (from the simulation) of the part  of the system of  binary mixture.
	$ARNPs$ (black particles) and passive particles (red particles). The two parameters  $S=4$ and $v=5\times10^4$.}
	%typical simulation snapshot of the model picture. {\color{red}Kindly mention the details like activity and size ratio}}
\label{fig:1}
\end{figure}
 
We  characterise the dynamics of both types of particles in the mixture for different system parameters (size ratio
$S$ and activity ${v}$). We first calculate the displacement of 
$ARNP$, and passive particles and  calculate their mean-square displacement $(MSD)$;
$\Delta_{a,p} (t) = \large \langle |{\bf r}(t+t_0)-{\bf r}(t_0)|^2\large \rangle$. 
The subscript $a$ and $p$ resemble the 
active and passive particles respectively.  $<..>$,  implies average over different reference 
times $t_0$'s, for all the particles of respective types and over $50$ independent realizations.

\begin{figure}
    \centering
    
    \includegraphics[width=1.0 \linewidth]{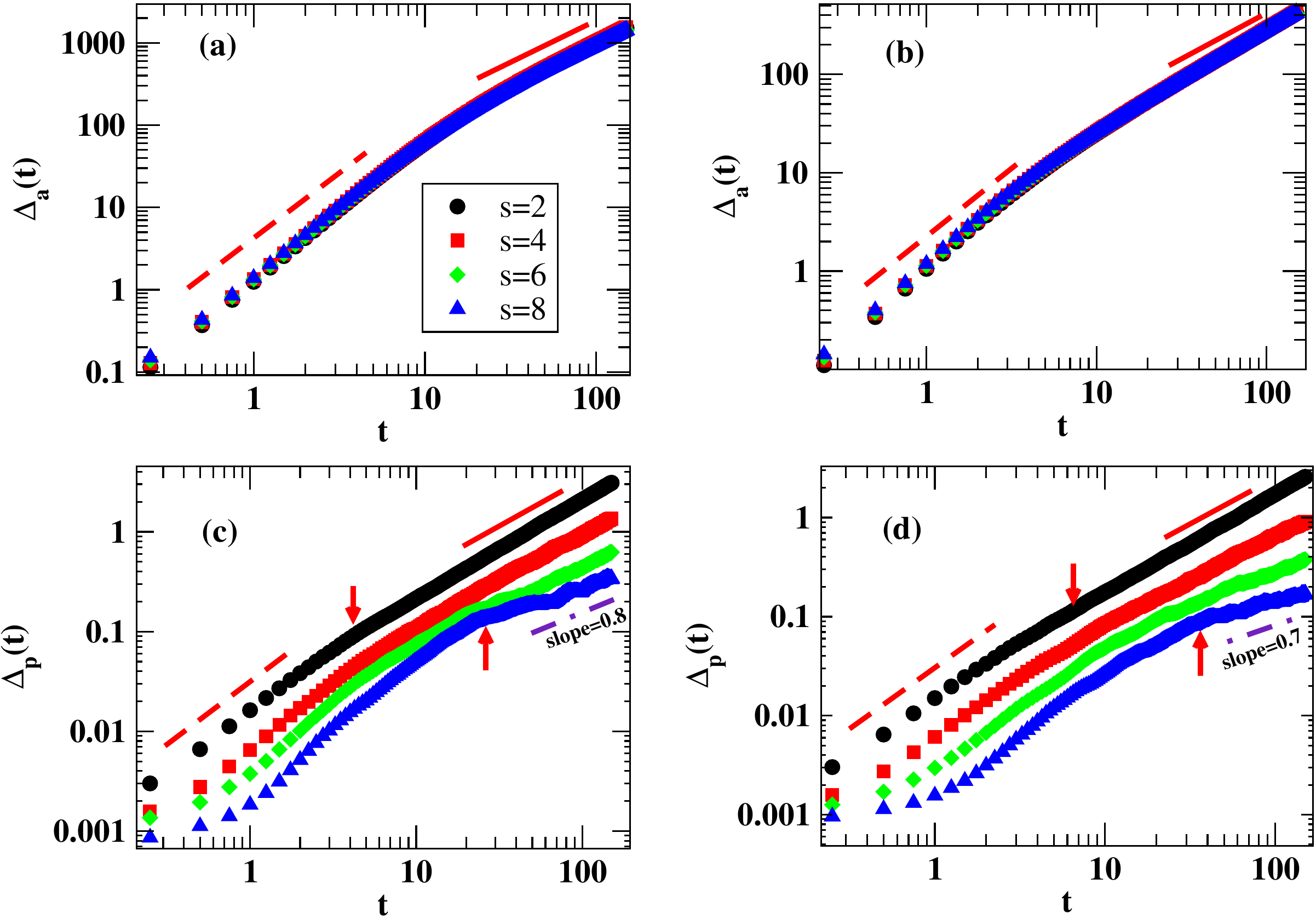}
	\caption{ (color online) Plot $\Delta_{a}(t)$ vs. $t$ for active (a-b) and passive particles $\Delta_{p}(t)$ (c-d), for  activity $v$= $5\times10^4, 1.2\times10^4$ with variation of different size ratios $S$. Dashed and solid lins are lines with slope $2$ and $1$ respectively. The two vertical arrowd in (c-d) shows the typical crossover time of $\Delta_{p}(t)$. The dotted dashed line in (c) and (d) is of slope $0.8$ and $0.7$ respectively.}
    \label{fig:2}
\end{figure}

\begin{figure}
   \centering
   \includegraphics[width=1 \linewidth] {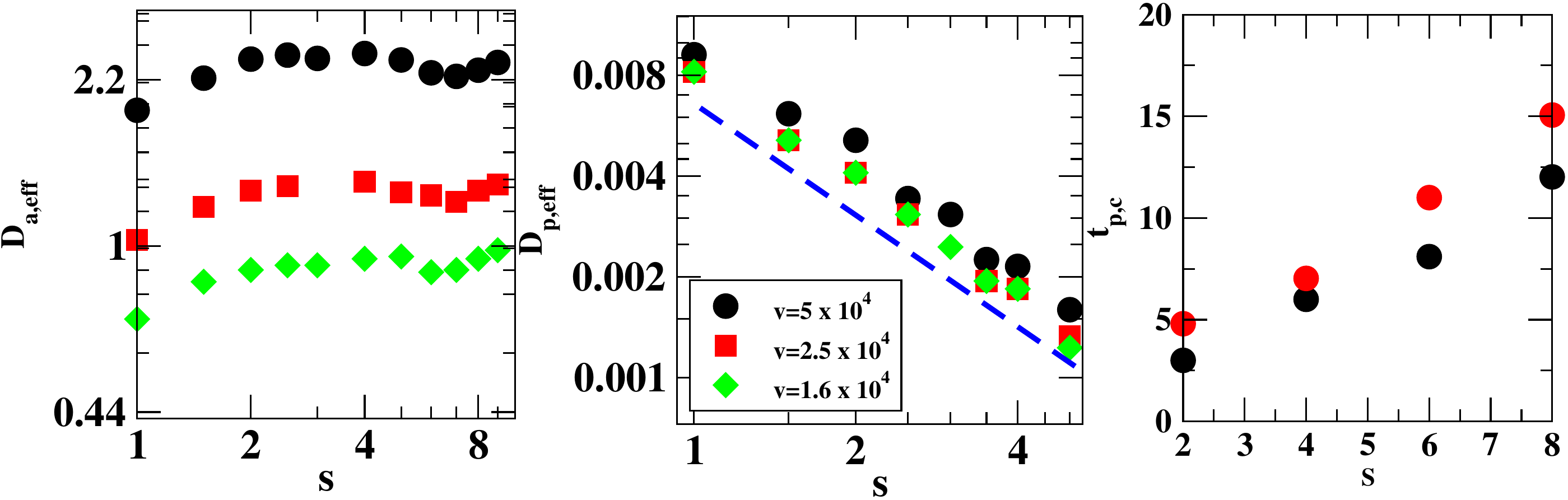}
    
	\caption{(color online) Plot shows variation of $D_{a,eff}$ with size ratio $S$ for different  activity $v$ for active particles (a), $D_{p,eff}$ vs. $S$ for passive particles (b), $t_{p,c}$  for passive particles  with size ratios $S$ for two different $v$=$5\times10^4,1.6\times10^4$(c). (Error of the order of symbol size).}
    \label{fig:3}  
\end{figure}

The Fig. \ref{fig:2}(a-b) shows the plot of  $MSD$ of $ARNPs$, $\Delta_a(t)$ for different size ratio S=2,4,6, and 8 keeping fixed
activities ${v}=5\times10^4$ and ${v}=1.2\times10^4$ respectively. We find that $\Delta_a(t)$ is  independent of the size 
ratio $S$ in all the cases and  shows an early time superdiffusive to late time diffusive dynamics.
In general the particle  follows the persistent random walk (PRW) and $MSD$ can be approximated as - 
\begin{equation}
	\Delta({t}) = 2d{D_{eff}}t[1-\exp(\frac{-t}{t_c})]
\label{eq:4}
\end{equation}
where $D_{eff}$ is the effective diffusivity in the steady state and $t_c$ is the typical crossover time from superdiffusion to 
diffusion. The effective diffusivity  of $ARNPs$, $D_{a, eff}$ shows weak dependence  with  size as shown in Fig. 
\ref{fig:3}(a). 
In Fig. \ref{fig:2}(c-d) we show the plot of 
$MSD$ of passive particles $\Delta_p(t)$ for different size ratio. 
%For larger $S$, $\Delta_p(t)$ shows an early time superdiffusive dynamics and late time subdiffusive behavior. 
%The two vertical arrows in Fig. \ref{fig:2}(c-d) indicates the typical crossover 
%time from early time superdiffusive to late time subdiffusive behavior. 
The dashed and solid lines are lines with slope 
$2$ and $1$ respectively 
%{\color{red} solid line creates confusion since it is for slope 1 whereas for subdiffusion regime slope should be less than 1.}. 
For  small size ratio, the $MSD$ is diffusive at the late time and slowly becomes subdiffusive for large $S$. The 
typical crossover time $t_{p,c}$ (as marked by two vertical arrows at the top and bottom curves).
The crossover time $t_{p,c}$ increases linearly with $S$ for both activities as shown in Fig.\ref{fig:3}(c).
Hence bigger passive particles spend more time in superdiffsuion. 
We further calculated the dependence of effective diffusivity of passive particles, $D_{p,eff}$ on size ratio $S$. 
The $D_{p,eff}$  decreases inversely as 
a function of  size ratio  as shown in Fig. \ref{fig:3}(b). It matches with the earlier results for the diffusion of 
Brownian disk moving in the equilibrium Stokes fluid \cite{stoke}.\\

\begin{figure}[hbt]
{\includegraphics[width=1.0 \linewidth] {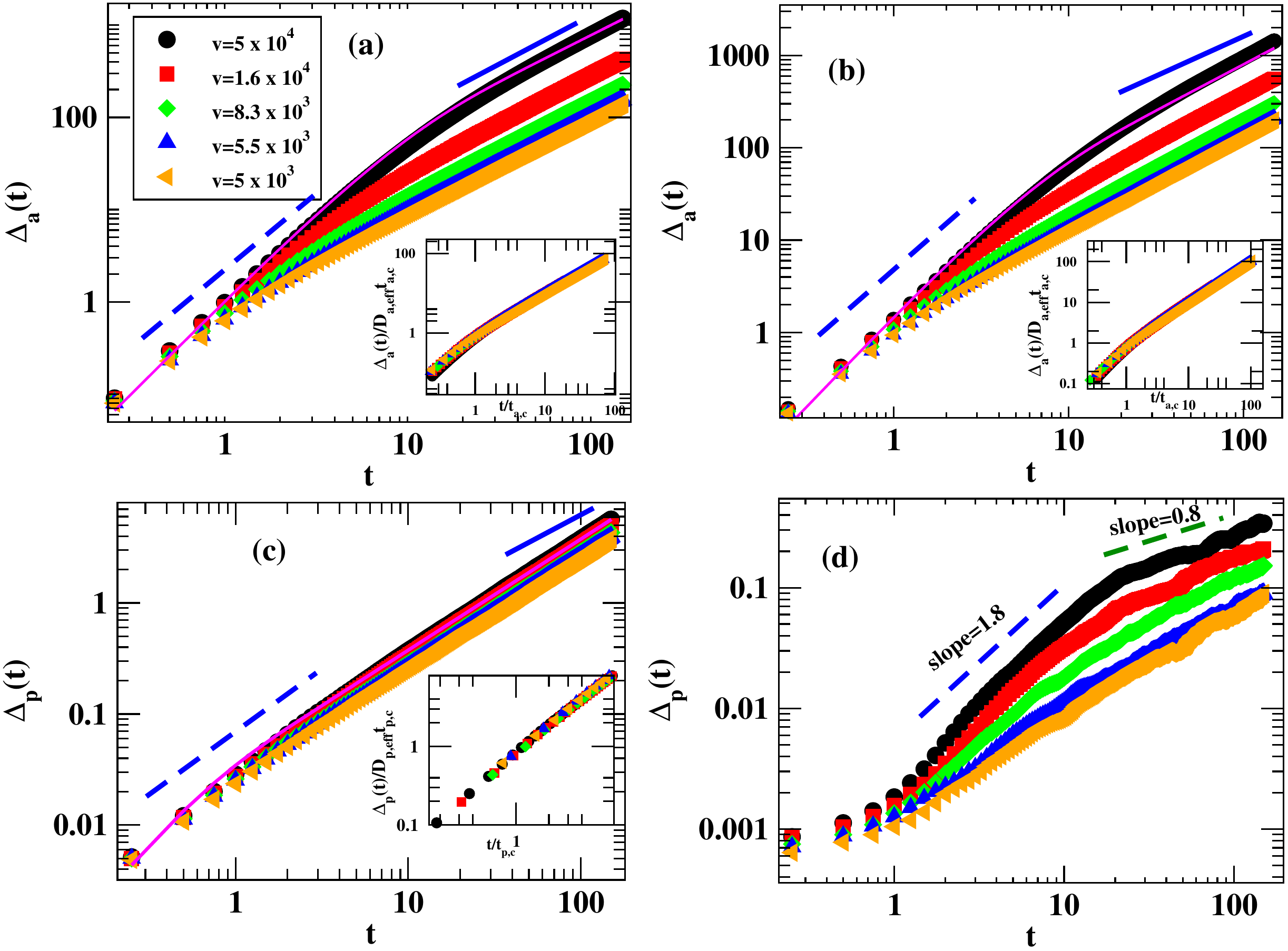}}
       \caption{(color online) The Plots (a)-(d) show the $\Delta(t)$ vs. $t$ for active $\Delta_{a}(t)$ (a-b) and 
	$\Delta_{p}(t)$ passive particles (c-d) for $S=$ $1,8$ with  $v$ (inset shows the scaling plot of $MSD$). 
	lines are fitting  function  using Eqn. {\ref{eq:4}}. 
	solid and dashed lines in (a-c) are of slope $2$ and $1$ respectively. The dashed and dotted dashed 
	lines in (d) are of slope $1.8$ and $0.8$ repectively.}
       \label{fig:4}
\end{figure}

\begin{figure}
    \centering
    \includegraphics[width=0.6 \linewidth]{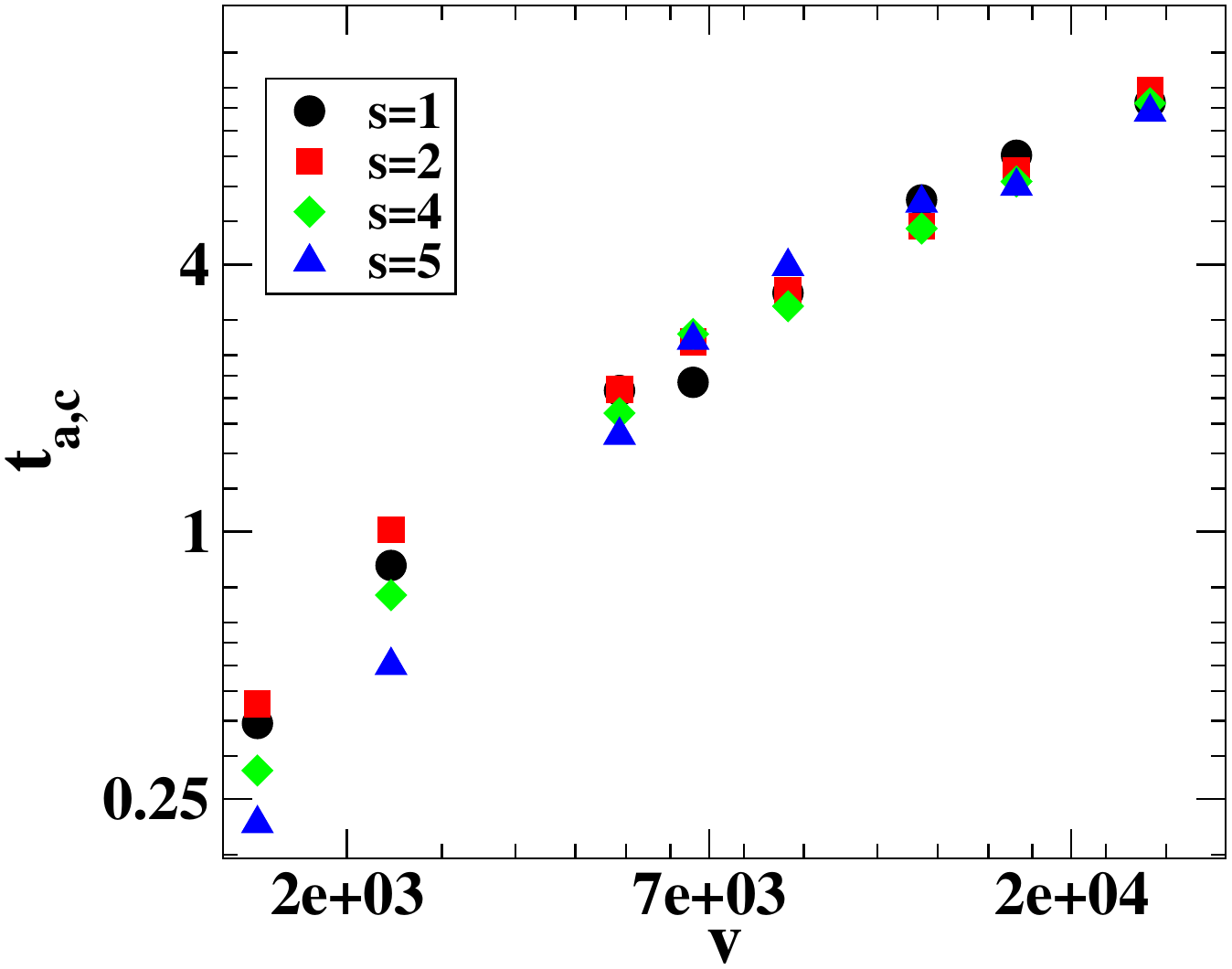}
	\caption{(color online) (a) Plot of  variation of  $t_{a,c}$ with  $v$ for different size ratios $S$.}
    \label{fig:5}
\end{figure}

We  also explored the system for  fixed  size ratios $S=1$ and $S=8$ and 
varying the activity $v$. For all activities the $ARNPs$ show the persistent 
random walk $(PRW)$ as given in Eq. \ref{eq:4} 
and
$MSD$ shows a crossover from early time ballistic to the late time diffusive behaviour. In Fig. \ref{fig:4}(a-b) we 
shows the plot of $MSD$ of $ARNPs$ particles, $\Delta_a(t)$ for different $v$ and fixed size ratios $S=1$ and $8$
respectively. The data points from the numerical simulation and lines are fit from the expression of $MSD$ as given in Eq. \ref{eq:4} for $v=5\times10^4$.
%S and $v=5\times10^4$
%are the fitting function as given in Eq. \ref{eq:4} for the
%PRW}. 
%The data points are data from the numerical simulation and lines.
The crossover time 
$t_{a,c}$, increases by increasing activity. The $t_{a,c}$ is obtained by fitting the $MSD$ $\Delta_a(t)$ of $ARNPs$ 
with the expression of $PRW$ as given in Eq. \ref{eq:4}.
In Fig. \ref{fig:5} we plot the crossover $t_{a,c}$ vs. $v$.
%by fitting 
%the $MSD$ with the expression as given in Eq. \ref{eq:4}. 
The $t_{a,c}$ increases 
linearly with increasing $v$ as shown in Fig. \ref{fig:5}.
We also calculated the $D_{a, eff}$, obtained from
the fitting. The variation of $D_{a, eff}$ with actvity will be discussed later. \\
   We show
the scaling collapse of $MSD$ by plotting on the $x-$axis the scaled time $t/t_{a,c}$ and $y-$axis the scaled 
$MSD$, $\frac{\Delta_{a}(t)}{D_{a, eff} t_{a,c}}$. 
We find  scaling collapse of data for both size ratios and for 
all activities as shown in the inset of Fig. \ref{fig:4}(a-b).

We also calculated the $MSD$ of passive particles $\Delta_p(t)$ for different actvities and for the two size ratios $S=1$ and $S=8$ as shown in Fig. \ref{fig:4}(c-d).
For small activity $S=1$, the passive particles also show a crossover from early time ballistic to late time
diffusive behavior and fitted well with the expression for the PRW as given in Eq. \ref{eq:4}.
Data shows a scaling collapse when  we plot the scaled time $t/t_{p,c}$ vs. scaled $MSD$, $\frac{\Delta_{p}}{D_{p, eff} t_{p,c}}$ \ref{fig:4}(c).
For large size ratio, $S=8$, passive particles show an early time ballistic but late time subdiffusion 
as we can see in the
 Fig. \ref{fig:4}(d). Hence the $MSD$ can not no longer be compared with the $PRW$.

\begin{figure} [hbt]
{\includegraphics[width=1.0 \linewidth]{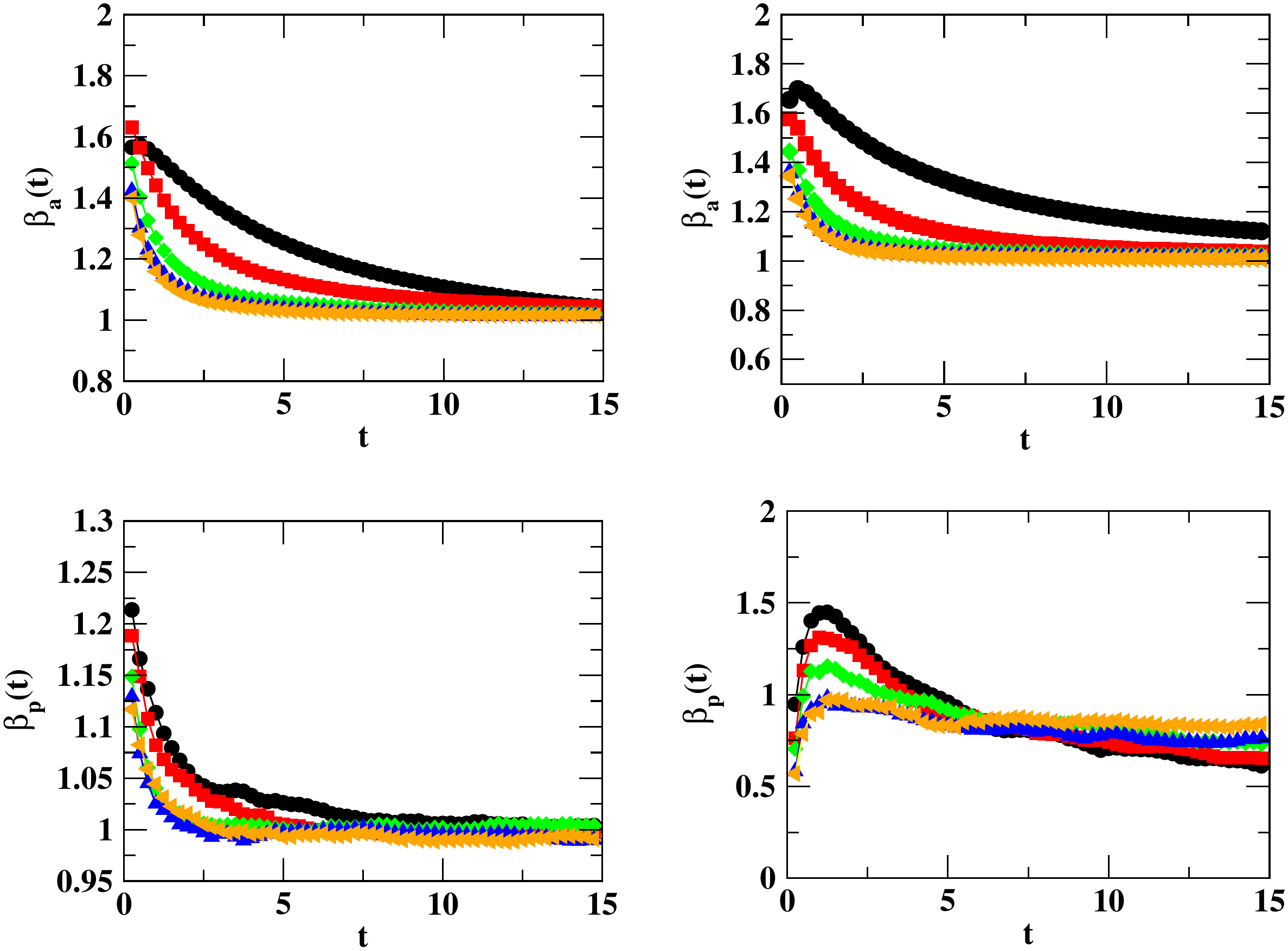}}
	\caption{(color online) In these plot we shows the $\beta(t)$ vs $t$, for active $\beta_{a}(t)$ (a-b) and passive $\beta_{p}(t)$ (c-d). Other parameters are the same as in Fig.{\ref{fig:4}}.}
\label{fig:6}
\end{figure}
We also investigated the dynamics of particles by
extracting the dynamic $MSD$ exponent $\beta(t)$ defined by $\Delta(t)\sim t^{\beta(t)}$, hence

\begin{equation}
\beta (t)= \frac{log_{10}[\Delta (10t)]}{log_{10}[\Delta(t)]}
\end{equation}

The Fig.{\ref{fig:6} is plotted for the same parameter as in \ref{fig:4}}. For the  $ARNPs$ $\beta_{(a)}(t)$ shows crossover from 
superdiffusive $\beta_{a}(t) > 1$ to diffusive $\beta_a(t)\sim 1$ regime. For the  passive particles for $S=1$, early time dynamics is  superdiffusive $\beta_{p}(t) > 1$ and becomes  diffusive $\beta_{p} \sim 1$ at late time.
Whereas for the  size ratio $8$ passive particles show superdiffusion $\beta_p(t) >1$ to 
subdiffusive $\beta_p <1$ motion. 
\subsection{Diffusivity and effective temperature of active particles in the mixture}
In order to further explore the concept of effective temperature \citep{temperature,temperature2,temperature3,temperature4} of the medium.   
%It  brings discussion about the validity of an effective equilibirum in this active mixture. 
Assuming an effective equilibirum, a  relation between 
an effective temperature (calculated from the speed distribution) and 
effective diffusivity calculated from $MSD$ of active particles can be written as;
$T_{a,eff}(\Delta) = D_{a,eff} /k_B$,
where $k_B$ is a constant  factor used as the fitting parameters.
 
We  calculated the speed distribution $P(s)$ of $ARNPs$. The particle
speed distributions determine the mean kinetic energy of the
particles. If the distribution follows a Maxwell-Boltzmann $(MB)$ form 
as always the case in fluids at equilibrium, the mean kinetic
energy is related to the thermodynamic temperature via the
equipartition theorem \cite{temperature}. 
We calculate the $p(s)$ and it follows the $MB$ distribution for different parameters.
Compariing it with standard $MB$ distribution we calculated the effective temperature $T_{a, eff}(v)$ as a function of 
activity $v$ for different size ratio $S$. 

\begin{figure} [hbt]
{\includegraphics[width=0.7 \linewidth]{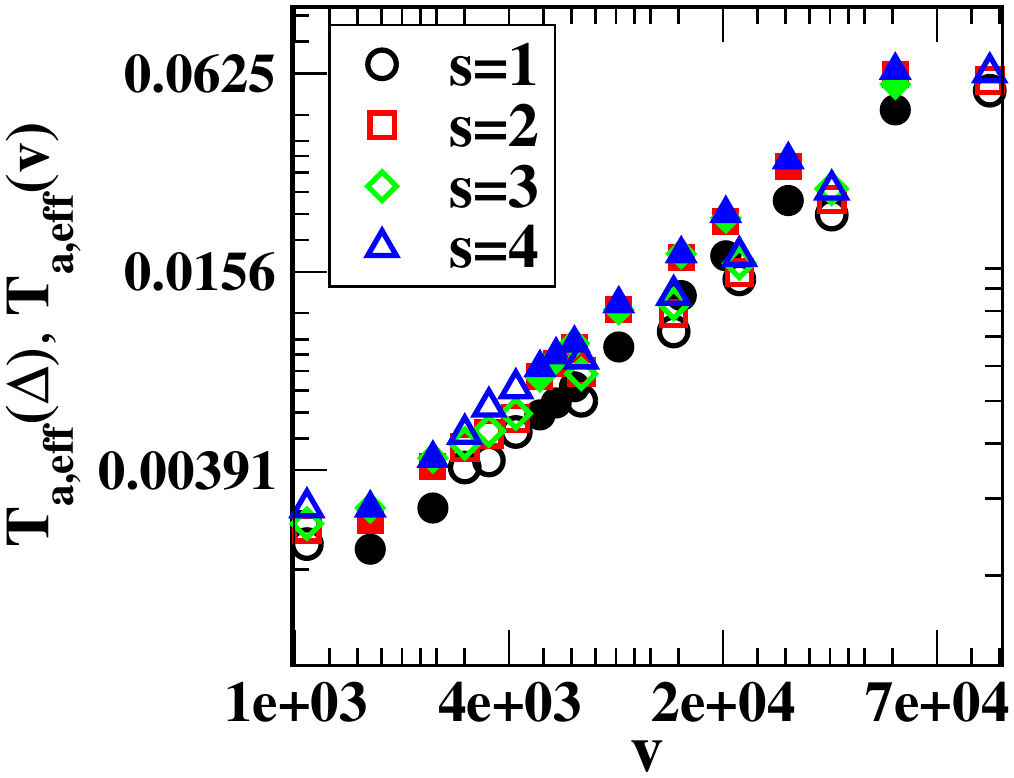}}
	\caption{(color online) $T_{a, eff}(\Delta)$ (open symbol) and $T_{a,eff}(v)$ (solid symbol) vs. $v$ 
	for different size ratios $S$. }
\label{fig:7}
\end{figure}

In Fig. \ref{fig:7}
we plot the variation of   $T_{a,eff}(v)$ and $T_{a, eff}(\Delta)$ vs. $v$ for different $S$. 
The data shows good match of both the effective temepartures. 
In active run and tumble particle system \cite{taillur}
$D_{eff}=v_{0}^{2}/d\lambda = v_{0} v r_{a}/d $. Where $d$ is the dimensionality of space. 
Hence $T_{a, eff}(\Delta)$ varies linearly with $v$ 
as shown in Fig. \ref{fig:7}.

{\section{Discussion}\label{secdis}}
We studied the dynamics of a binary mixture of disk-shaped active run and tumble and passive particles 
on a two-dimensional substrate. Both types of particles are athermal in nature.
The activity of active particles is controlled by their tumbling rate.  The size of $ARNPs$ is fixed whereas it is varied for passive particles. Further, in the  mixture of $ARNPs$  particles, the  $MSD$ show the early time ballistic behavior 
and late time diffusive motion with increasing value of $v$ and size ratio $S$. 
The passive particles show the crossover from late time subdiffusive to diffusive dynamics
on increasing $v$ and decreasing $S$. The late time effective diffusivity of passive particles
$D_{p, eff}$ decay monotonically with their size as found in equilibrium passive Stokes fluid \cite{stoke}.
The effective diffusivity of $ARNPs$ increases linearly with their activity and shows a good match with the 
effective temperature obtained from the steady-state speed distribution with the Maxwell-Boltzmann
distribution. \\
Hence our study explores dynamics and steady-state of $ARNPs$ and passive particles in the mixture and shows an
effective equilibrium in the system.
Our particle-size dependence of  $MSD$ of passive particles 
in the presence of active run and tumble particles 
has important applications in particle sorting in  different types of 
fluids like-microfluidic devices \cite{drug}.

\section{acknowledgement}
We thank I.I.T. (BHU) Varanasi computational fascility. Vivek Semwal thanks
DST INSPIRE (INDIA) for the research fellowship. S. Mishra thanks DST, SERB (INDIA), Project No. ECR/2017/000659 for partial financial support.

%\end{references}
\end{document}